\documentclass[9pt,twocolumn,twoside]{osajnl}
\journal{josab} % Choose journal (ao, aop, josaa, josab, ol, optica, pr)

\setboolean{shortarticle}{false}
\usepackage{amsmath}
\usepackage{bibentry}
\usepackage[utf8]{inputenc}
\usepackage{tikz}
\usepackage{svg}
\usepackage{comment}
\usepackage{float}
\usetikzlibrary{matrix,shapes,arrows,positioning,chains}
\newcommand{\ignore}[1]{}
\newcommand{\nobibentry}[1]{{\let\nocite\ignore\bibentry{#1}}}

\newcommand{\da}[1]{\text{d}#1 }

\graphicspath{{figs_josab/}}
\title{Numerical Simulation of Noise in Pulsed Brillouin Scattering}

\author[1,*]{Oscar A. Nieves}
\author[1]{Matthew D. Arnold}
\author[2]{Michael. J. Steel}
\author[2]{Miko\l{}aj K. Schmidt}
\author[1]{Christopher G. Poulton}

\affil[1]{School of Mathematical and Physical Sciences, University of Technology Sydney, 15 Broadway, Ultimo NSW 2007, Australia.}
\affil[2]{Macquarie University Research Centre for Quantum Engineering (MQCQE), MQ Photonics Research Centre, Department of Physics and Astronomy, Macquarie University, NSW 2109, Australia.}

\affil[*]{Corresponding author: oscar.a.nievesgonzalez@student.uts.edu.au}

\begin{abstract}
 We present a numerical method for modelling noise in Stimulated Brillouin Scattering (SBS). The model applies to dynamic cases such as optical pulses, and accounts for both thermal noise and phase noise from the input lasers. Using this model, we compute the statistical properties of the optical and acoustic power in the pulsed spontaneous and stimulated Brillouin cases, and investigate the effects of gain and pulse width on noise levels. We find that thermal noise plays an important role in the statistical properties of the fields, and that laser phase noise impacts the SBS interaction when the laser coherence time is close to the time-scale of the optical pulses. This algorithm is applicable to arbitrary waveguide geometries and material properties, and thus presents a versatile way of performing noise-based SBS numerical simulations, which are important in signal processing, sensing, microwave photonics and opto-acoustic memory storage.
\end{abstract}

\setboolean{displaycopyright}{true}

\begin{document}

\maketitle

\section{Introduction}
Stimulated Brillouin Scattering (SBS) is an opto-acoustic process that results from the interaction between two counter-propagating optical fields, the pump and the Stokes, as well as an acoustic wave inside a dielectric medium~\cite{eggleton2019, pant2014, brillouin1922, boyd2003,kobyakov2010}. This interaction has been used for applications including narrow-band radio-frequency (RF) and optical signal filtering~\cite{choudhary2018, jiang2018}, phase conjugation and precision spectroscopy~\cite{eggleton2019}, novel laser sources~\cite{stein2007, loh2015optica}, and in recent experiments in opto-acoustic memory storage~\cite{merklein2017}. One of the key challenges of simulating the SBS interaction is modelling of thermal noise, which is present in all real systems and which can significantly affect performance~\cite{boyd1990,gaeta1991,ferreira1994}. Simulating noise in the SBS equations is complicated because of the nonlinear coupling between the envelope fields: beyond the undepleted pump regime the noise is  multiplicative and can only be understood in the context of statistical moments using multiple independent realizations~\cite{nieves2021}. Thermal noise in SBS has been simulated numerically in earlier studies~\cite{boyd1990,gaeta1991}, with these investigations concentrating on the noise properties of the Stokes signal that
arises spontaneously in response to a strong, continuous-wave (CW) pump. More recent simulations~\cite{behunin2018} have incorporated both thermal and laser noise in the SBS interaction, but have focused on single-mode structures such as micro-ring resonators in steady-state laser conditions. A numerical method for solving the transient SBS equations with laser and thermal noise is needed for accurately predicting and characterizing the noise in modern integrated SBS waveguide experiments~\cite{zhang2011, pant2014, merklein2017}.

In this paper, we present a numerical method by which the transient SBS equations with thermal noise can be solved for pulses of arbitrary shape and size, in arbitrary waveguide geometries. The method allows for the inclusion of input laser noise in the form of stochastic boundary conditions. We apply this method to the case of a short chalcogenide waveguide and use the model to compute the statistics of the output envelope fields. We examine the dynamics of the noise 
when the Stokes arises spontaneously from the thermal field, and for the case when it is seeded with an input pulse at the far end of the waveguide. We demonstrate the transition from 
the low-gain, short pulse case, in which noise is amplified by the pump, to the high gain, long pulse regime in which coherent amplification occurs. In this latter situation, we show that
while the output pulses remain smooth, significant fluctuations in the peak powers arising from the thermal field can persist.
We also show that, within the framework of this model, phase noise from the pump only has a significant impact on Stokes noise when the laser coherence time matches the time scales of the pulses involved in the interaction.
 Finally, we investigate the convergence of this numerical method, and find that it yields linear convergence in both the average power and variance of the power for three fields in the SBS interaction, which is in agreement with the Euler-Mayurama scheme for solving stochastic ordinary differential equations.

%%%%%%%%%%%%%%%%%%%%%%%%%%%%%%%%%%%%%%%%%%%%%%%%%%%%
% METHOD
%%%%%%%%%%%%%%%%%%%%%%%%%%%%%%%%%%%%%%%%%%%%%%%%%%%%
\section{Method}
\subsection{The SBS equations}
\begin{figure}[ht]
\centering
\includegraphics[width=0.5\textwidth]{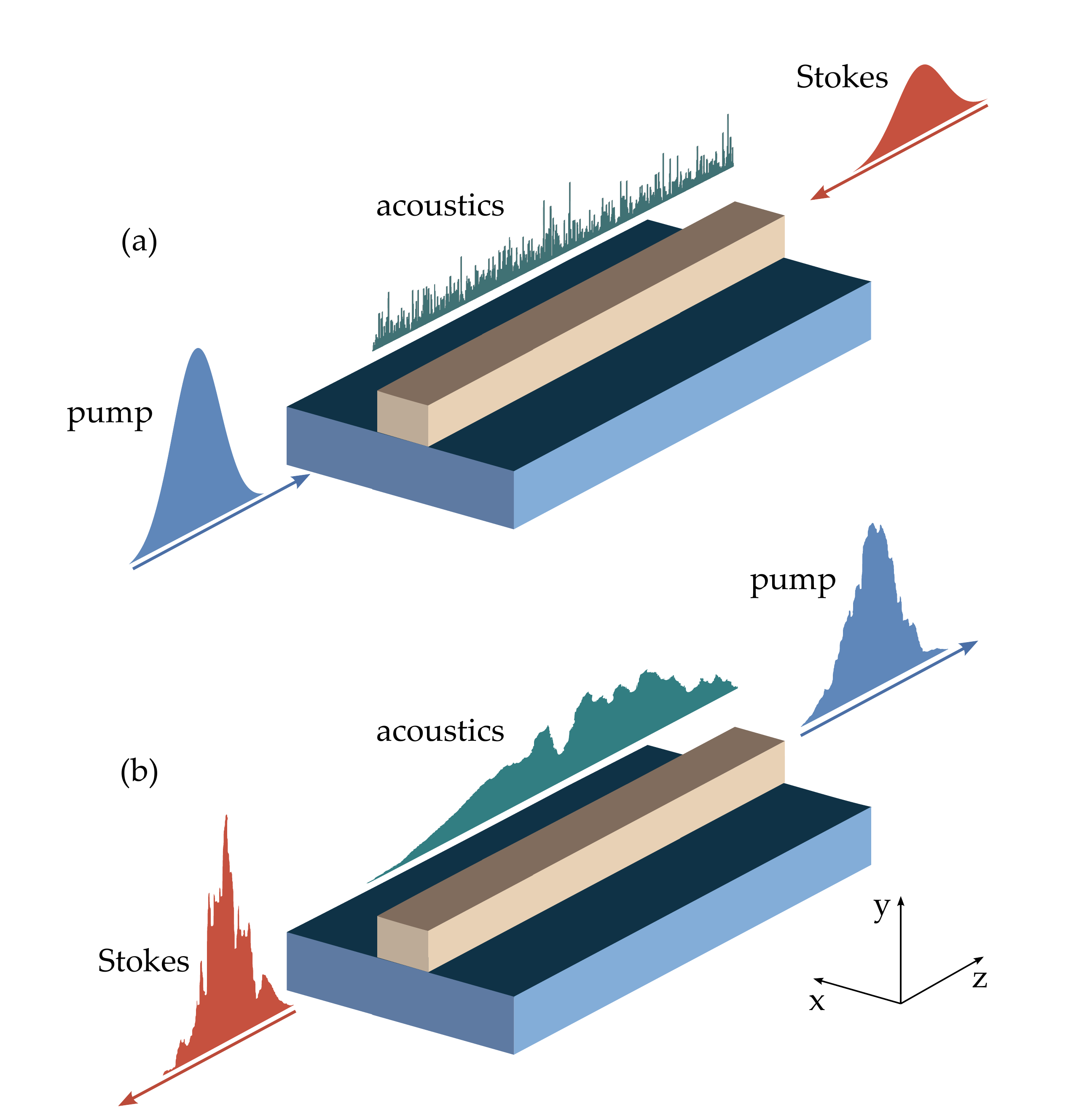}
\caption{Illustration of the SBS interaction, showing the pump, Stokes and acoustic powers on a photonic chip waveguide. In (a), the pump and Stokes pulses are injected into opposite ends of the waveguide, and the acoustic field is made up of random thermal fluctuations. In (b), the optical fields have interacted inside the waveguide, the Stokes depletes the pump to gain some energy, and the rest of the energy goes to the acoustic field.}
\label{fig:Numerical_Figure_1}
\end{figure}

We consider backward SBS interactions in a waveguide of finite length $L$ along the $z$-axis, in which a pump pulse with angular frequency $\omega_1$ is injected into the waveguide at $z=0$ and propagates in the positive $z$-direction, while a signal pulse is injected at $z=L$ and propagates in the negative $z$-direction, as shown in Fig.~\ref{fig:Numerical_Figure_1}. The spectrum of the signal pulse
is  centered around the Brillouin Stokes frequency $\omega_2=\omega_1-\Omega$, which is down-shifted from the pump by the Brillouin shift $\Omega$, and its spectral extent lies entirely within the Brillouin linewidth $\Delta\nu_B$.
 When these two 
pulses interact, energy is transferred from the pump to the signal via the acoustic field, resulting in
coherent amplification of the signal around the Brillouin frequency. At the same time, as the pump moves through the waveguide, it interacts with the thermal phonon field and generates an incoherent contribution to the Stokes field which also propagates in the negative $z$-direction. This noisy Stokes field combines with the coherent signal to form a noisy amplified output field centered around the Stokes frequency. The interaction can be modelled using three envelope fields
for the pump ($a_1(z,t)$), Stokes ($a_2(z,t)$) and acoustic field ($b(z,t)$),
according to the equations \cite{nieves2021}
\begin{subequations}
\begin{align}
\label{eq:A1pde}\frac{\partial a_1}{\partial z} + \frac{1}{v} \frac{\partial a_1}{\partial t}  + \frac{1}{2}\alpha a_1 &= i\omega_1 Q_1 a_2 b^*,\\
\label{eq:A2pde}\frac{\partial a_2}{\partial z} - \frac{1}{v} \frac{\partial a_2}{\partial t}  - \frac{1}{2}\alpha a_2 &=  i\omega_2 Q_2 a_1 b,\\
\frac{\partial b}{\partial z} + \frac{1}{v_a} \frac{\partial b}{\partial t}  + \frac{1}{2}\alpha_{\text{ac}} b &= i\Omega Q_a a_1^* a_2 + \sqrt{\sigma} R(z,t). \label{eq:bpde}
\end{align}
\end{subequations}

% --- FIG. Illustration of SBS

\noindent Here $\alpha$ and $\alpha_{\text{ac}}$ are the optical and acoustic loss coefficients respectively (in units of m$^{-1}$), along with optical group velocity $v>0$ and acoustic group velocity $v_a>0$. The envelope fields $a_{1,2}$ and $b$ have units of W$^{1/2}$. The coefficients $Q_{1,2,a}$ represent the coupling strength of the SBS interaction, which depend on the optical and acoustic modes of the waveguide~\cite{sturmberg2019}; from local conservation of energy, we have $Q_2 = Q_1^*$ and $Q_a=Q_1$~\cite{wolff2015stimulated}. Here we focus on the single acoustic-mode case, which we can choose by tuning the laser frequencies and relying on the large free spectral range of the acoustic modes. This model can further be extended by including additional acoustic fields with their own opto-acoustic coupling constants and potentially different noise properties~\cite{wolff2015stimulated}.

The boundary conditions for the pump and signal fields are applied by specifying the input values $a_1(0,t)$ and  $a_2(L,t)$ respectively. These boundary conditions depend on the laser properties, such as the linewidth, and may contain noise. Thermal noise in the waveguide is introduced through the complex-valued space-time white noise function $R(z,t)$, which has mean $\left\langle R(z,t)\right\rangle = 0$ and auto-correlation function $\left\langle R(z,t) R^*(z',t') \right\rangle = \delta(z-z')\delta(t-t')$. The noise strength is derived by analytically solving~\eqref{eq:bpde} in the absence of any optical fields~\cite{nieves2021}, and is  $\sigma = k_B T\alpha_{\text{ac}}$ where $k_B$ is the Boltzmann constant and $T$ is the temperature of the waveguide.

We begin with the observation that the propagation distance of the 
acoustic wave over the time-scale of the interaction is very small~\cite{boyd1990}. We therefore apply the limit $\partial_z b \rightarrow 0$ in~\eqref{eq:bpde}, which simplifies to
\begin{equation}\label{eq:bpde2}
\frac{1}{v_a} \frac{\partial b}{\partial t}  + \frac{1}{2}\alpha_{\text{ac}} b = i\Omega Q_a a_1^* a_2 + \sqrt{\sigma} R(z,t).
\end{equation}
This has the formal solution
\begin{equation}\label{eq:bsol}
b(z,t) = iv_a \Omega Q_a \int_{-\infty}^{t} e^{-\frac{\Gamma}{2}(t-s)} a_1^*(z,s)a_2(z,s)\da{s} + D(z,t),
\end{equation}
where $\Gamma=v_a\alpha_{\text{ac}}$ is decay rate of the acoustic field, namely $\Gamma = 1/\tau_a$, and is related to the Brillouin linewidth via $\Gamma = 2\pi\Delta\nu_B$. The thermal noise enters through the function
\begin{equation}\label{eq:Df}
D(z,t) = v_a\sqrt{\sigma} \int_{-\infty}^{t} \mathrm{e}^{-\frac{\Gamma }{2} (t-s)} R(z,s)\da{s}.
\end{equation}
This function $D$ is a stochastic integral with zero mean $\left\langle D(z,t) \right\rangle = 0$ since the function $R(z,s)$ is itself a zero-mean stochastic process. The auto-correlation function of $D$ at two times and two points in space is found by following the derivation in~\cite{nieves2021}, which uses the stochastic Fubini theorem~\cite{veraar2012} to obtain the expression:
\begin{equation}
\left\langle D(z,t) D^*(z',t')\right\rangle = \frac{v_a\sigma}{\alpha_{\text{ac}}}\delta(z-z') \exp\left\{-\frac{\Gamma}{2}\left|t - t'\right|\right\}.
\end{equation}
Upon substitution of~\eqref{eq:bsol} into~\eqref{eq:A1pde} and~\eqref{eq:A2pde}, and assuming that the fields $a_{1,2}$ are everywhere zero for $t<0$, we obtain the pair of equations
\begin{multline}\label{eq:a1_PDE}
\frac{\partial a_1}{\partial z} + \frac{1}{v}\frac{\partial a_1}{\partial t} + \frac{1}{2}\alpha a_1 = i\omega_1 Q_1 a_2(z,t) D^*(z,t) \\
-\frac{1}{4}g_1\Gamma a_2(z,t) \int_{0}^{t} e^{-\frac{\Gamma}{2}(t-s)} a_1(z,s)a_2^*(z,s)\da{s},
\end{multline}
\begin{multline}\label{eq:a2_PDE}
\frac{\partial a_2}{\partial z} - \frac{1}{v}\frac{\partial a_2}{\partial t} - \frac{1}{2}\alpha a_2 = i\omega_2 Q_2 a_1(z,t)  D(z,t)\\
-\frac{1}{4}g_2 \Gamma a_1(z,t) \int_{0}^{t} e^{-\frac{\Gamma}{2}(t-s)} a_1^*(z,s)a_2(z,s)\da{s},
\end{multline}
where $g_1 = g_0 \omega_1/\omega_2$, $g_2 = g_0$, and the SBS gain parameter
$g_0 = 4v_a\omega_2\Omega|Q_2|^2 / \Gamma$ (with units of m$^{-1}$W$^{-1}$)~\cite{nieves2021}. 

The approach of the numerical method is to solve
~\eqref{eq:a1_PDE} and~\eqref{eq:a2_PDE} in a stepwise manner to find the optical fields; the optical fields at each time step are then substituted into~\eqref{eq:bsol} to obtain the acoustic envelope field, and the process is repeated.
At each time step the solution requires calculation of the thermal noise function $D(z,t)$ which behaves as a random walk in time while remaining white in space.
The optical equations are solved with the input boundary conditions $a(0,t) = a_p(t)$ and $a(L,t)=a_s(t)$; in general, these boundary conditions may be stochastic to account for noise in the input lasers.
In the following we first describe the approach taken to compute the thermal noise function, then discuss the inclusion of noise into the boundary conditions, before describing the iterative algorithm itself.

It should be noted that it is also possible to solve~\eqref{eq:A1pde},~\eqref{eq:A2pde} and~\eqref{eq:bpde} directly without integrating the acoustic envelope field in time first (as in~\eqref{eq:bsol}), and this procedure would yield the same results. However, since the thermal background field is assumed to be in an equilibrium state by $t = 0$, this alternative method would require simulating the acoustic envelope field for a very long time $t<0$. This is computationally less efficient and poses no advantages over the present method.

\subsection{Computing the thermal noise function}
\begin{figure}[ht]
\centering
\includegraphics[width=0.5\textwidth]{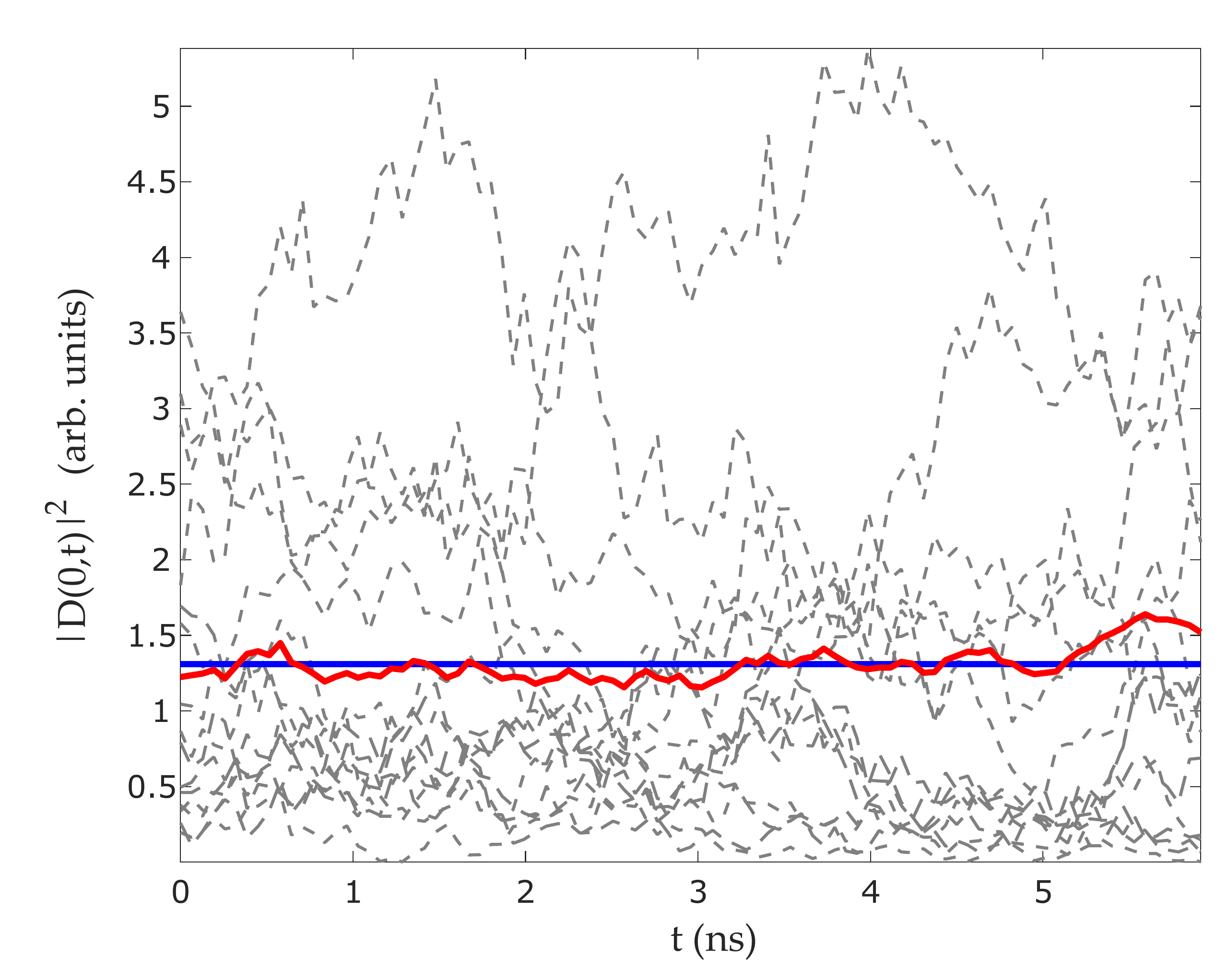}
\caption{Multiple independent realizations (dashed grey) of the modulus squared of the thermal function $|D(z_j,t)|^2$ at an arbitrary position $z_j$, the numerical ensemble average over 20 realizations (red) and the analytic ensemble average (blue). We use a temperature of $T=300$ K, $v_a=2500$ m/s, $\tau_a = 5.3$ ns, $\Delta z = 0.79$ mm and $\Delta t = 6.43$ ps.}
\label{fig:Dplots}
\end{figure}

The function $D(z,t)$ contains all the thermal noise information about the system. To model $D(z,t)$ numerically, we note that its evolution in time corresponds to an Ornstein-Uhlenbeck process~\cite{uhlenbeck1930}. Equation~(\ref{eq:Df}) can be written in It\^{o} differential form~\cite{van1976} as
\begin{equation}\label{eq:dD}
\da{D}(z_j,t) = -\frac{1}{2}\Gamma D(z_j,t)\da{t} + v_a\sqrt{\sigma} R(z_j,t)\da{t},
\end{equation}
where the $z$ axis is discretized on the equally spaced grid $z_j$ with spacing $\Delta z$. We know that $R(z_j,t_n)\da{t} =\frac{1}{\sqrt{\Delta z}}\da{W_j(t_n)}$ where $\da{W_j(t_n)}$ is the standard complex-valued Wiener increment in time, and the scaling factor arises from the Dirac-delta nature of the continuous-space auto-correlation function of $D(z,t)$. The complex increment $\da{W_j(t)}$ is a linear combination of two independent real Wiener processes
\begin{equation}
    \da{W_j(t)} = \frac{1}{\sqrt{2}}\left[\da{W_j^{(1)}(t)} + i \da{W_j^{(2)}(t)} \right],
\end{equation}
where $\left\langle \da{W_j^{(p)}(t)}\da{W_j^{(q)}(t)}\right\rangle = \delta_{pq} \da{t}$, where $\delta_{pq}$ is the Kronecker delta. Integrating~\eqref{eq:dD} from 0 to $t$ yields the analytic solution
\begin{equation}\label{eq:D_analytic}
    D(z_j,t) = e^{-\frac{1}{2}\Gamma t} D_0(z_j) + v_a\sqrt{\frac{\sigma}{\Delta z}} \int_{0}^{t} e^{-\frac{\Gamma}{2} (t-s)} \da{W_j(s)},
\end{equation}
where $D_0(z_j)$ is the cumulative random walk from $t=-\infty$ up to $t=0$. This quantity is calculated using
\begin{equation}\label{eq:D_0}
    D_0(z_j) = \frac{1}{\sqrt{2}}\left[\mathcal{N}_{z_j}^{(1)}\left(0, \frac{v_a \sigma}{\alpha_{\text{ac}} \Delta z }\right) + i\mathcal{N}_{z_j}^{(2)}\left(0, \frac{v_a \sigma}{\alpha_{\text{ac}} \Delta z }\right)  \right],
\end{equation}
where $\mathcal{N}_{z_j}^{(1,2)}\left(0, v_a \sigma/\alpha_{\text{ac}} \Delta z\right)$ are normal random variables with zero mean and variance $v_a\sigma/(\alpha_{\text{ac}}\Delta z)$, independently sampled at each $z_j$. Numerically, we can compute the integral in~\eqref{eq:D_analytic} following the procedure in Appendix A. Thus, we simulate~\eqref{eq:D_analytic} as a random walk using discrete increments $\Delta t$
\begin{multline}\label{eq:D2}
    D(z_j, t_{n+1}) = e^{-\frac{1}{2}\Gamma\Delta t}D(z_j, t_n)\ \\
    +\gamma(\Delta t) \left[\mathcal{N}_{z_j,t_n}^{(1)}\left(0, 1\right) + i\mathcal{N}_{z_j,t_n}^{(2)}\left(0,1\right)\right],
\end{multline}
where
\begin{equation}
\gamma(\Delta t) = v_a\sqrt{\frac{\sigma\left( 1-e^{-\Gamma\Delta t} \right)}{2\Delta z\Gamma}},
\end{equation}
and setting the initial value as $D(z_j,t_0) = D_0(z_j)$. The random numbers $\mathcal{N}_{z_j,t_n}^{(1,2)}\left(0, 1\right)$ are independently sampled at each point $(z_j,t_n)$. Figure~\ref{fig:Dplots} shows multiple realizations of $D(z_j,t)$ at an arbitrary point $z_j$ and its ensemble average.

% --- FIG. Random Walks D

%%% --- BOUNDARY CONDITIONS / LASER NOISE --- %%%
\subsection{Noisy boundary conditions}
\begin{figure}[ht]
\centering
\includegraphics[width=0.5\textwidth]{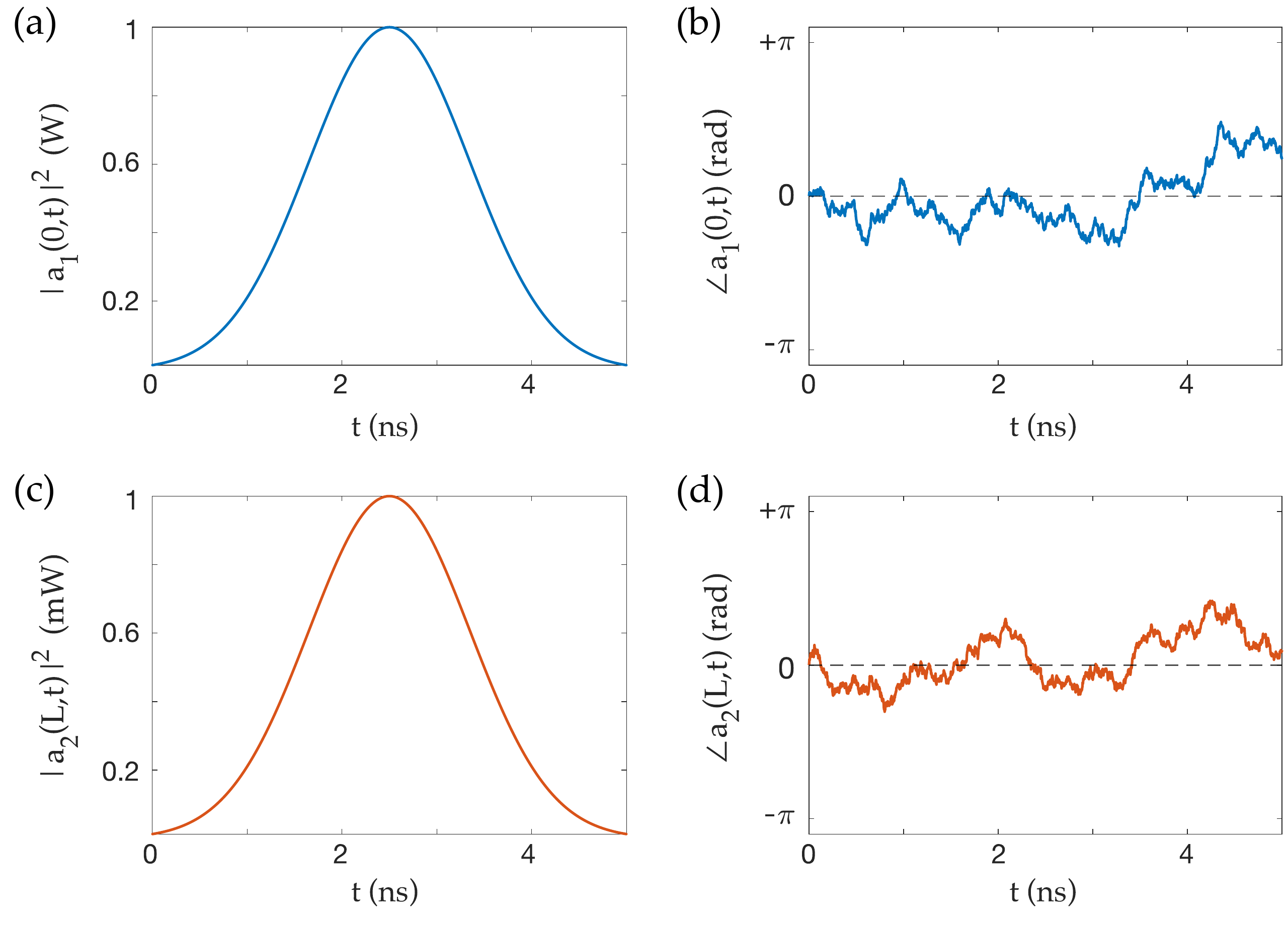}
\caption{Single realization of the noisy boundary conditions. The plots show (a) pump power, (b) pump phase, (c) Stokes power and (d) Stokes phase. Both pulses are Gaussian with FWHM of 2 ns. The laser linewidth used here is $\Delta\nu_{L} = 100$ MHz.}
\label{fig:phase_noise}
\end{figure}

Input laser noise can be an important feature in SBS experiments. In the context of the SBS envelope equations, it enters in the form of random phase fluctuations at the inputs of the waveguide, namely $z=0$ for the pump field and $z=L$ for the Stokes field. We simulate this laser phase noise in the input fields by modeling the boundary conditions as
\begin{align}
    \label{eq:ap} a_1(0,t_n) &= a_p(t_n) = \sqrt{P_1^{\text{in}}(t_n) } e^{i\phi_1(t_n)},\\
    \label{eq:as} a_2(L,t_n) &= a_s(t_n) = \sqrt{P_2^{\text{in}}(t_n) }  e^{i\phi_2(t_n)},
\end{align}
where $P_1^{\text{in}}(t)$ and $P_2^{\text{in}}(t)$ are deterministic envelope shape functions for the pump and Stokes fields respectively representing input power from the lasers. The variables $\phi_1(t)$ and $\phi_2(t)$ are stochastic phase functions modeled as zero-mean independent Brownian motions. The variation in the phase $\phi(t)$ is related to the laser's intrinsic linewidth $\Delta\nu_L$, or conversely the coherence time $\tau_{\text{coh}} = 1/(\pi\Delta\nu_L)$, via the expression $\left\langle \left[\phi(t+\tau)-\phi(t)\right]^2\right\rangle = 2\pi\Delta\nu_L |\tau|$, where $\tau=t'-t$ for the two times $t'$ and $t$~\cite{moslehi1986, debut2000, wei2012}. Following a similar numerical procedure to~\cite{atzmon2009}, we compute $\phi_j(t)$ as
\begin{equation}
\phi_j(t) =\sqrt{2\pi \Delta\nu_{L}} \int_{0}^{t} \da{W_j(s)},
\end{equation}
where $\da{W(s)}$ is a real-valued Wiener process increment in time. To generate the random walk numerically, we cast this integral as an It\^{o} differential equation $\da{\phi_j(t)} = \sqrt{2\pi\Delta \nu_L} \da{W_j(t)}$, which is discretized using an Euler-Mayurama~\cite{kloeden1992} scheme as
\begin{equation}\label{eq:phi_noise}
\phi_j(t_{n+1}) = \phi_j(t_n) + \sqrt{2\pi\Delta \nu_L} \sqrt{\Delta t}\ \mathcal{N}_{t_n}(0,1),
\end{equation}
where $\mathcal{N}_{t_n}(0,1)$ is a standard normally distributed random number sampled at each $t_n$. A simulation of a single realization of the noisy boundary conditions is shown in Figure~\ref{fig:phase_noise}.

% --- FIG 2 - Phase Noise

%%% --- SPLIT STEP METHOD --- %%%
\subsection{The numerical algorithm}
We now present the main numerical algorithm of this paper. The algorithm consists of two consecutive steps: first, we solve~Eqs.~(\ref{eq:A1pde}) and~(\ref{eq:A2pde}) in the absence of optical loss or nonlinear interactions. In other words, we solve the following pair of advection equations
\begin{align}
    \frac{\partial a_1}{\partial z} + \frac{1}{v} \frac{\partial a_1}{\partial t} &= 0,\\
    \frac{\partial a_2}{\partial z} - \frac{1}{v} \frac{\partial a_2}{\partial t} &= 0.
\end{align}
With the boundary conditions $a_1(0,t) = a_p(t)$ and $a_2(L,t) = a_s(t)$, these have the elementary solutions
\begin{align}
    \label{eq:a1_shift_1} a_1(z,t) &= a_p\left(t - \frac{z}{v}\right),\\
    \label{eq:a2_shift_1} a_2(z,t) &= a_s\left(t - \frac{L-z}{v}\right).
\end{align}
Setting the numerical grid parameter $\Delta z = v\Delta t$ further simplifies~\eqref{eq:a1_shift_1} and~\ref{eq:a2_shift_1} to
\begin{align}
    \label{eq:A1shift} a_1(z_j,t_n) \ &\leftarrow \ a_1(z_{j-1},t_{n-1}),\\
    \label{eq:A2shift} a_2(z_j,t_n) \ &\leftarrow \ a_2(z_{j+1},t_{n-1}),
\end{align}
such that the optical fields are shifted in space by exactly $\Delta z$ during each time iteration. The envelope field $b(z,t)$ is assumed to remain stationary in space during each time step, as is typical in the context of SBS experiments involving pulses~\cite{nieves2021}. After the fields are shifted across the waveguide, we solve the time evolution equations at each point $z_j$ independently; i.e. we solve
\begin{multline}
\frac{1}{v}\frac{\partial a_1(z_j,t)}{\partial t} =-\frac{1}{2}\alpha a_1(z_j,t) - \frac{1}{4}g_1\Gamma a_2(z_j,t) I_{1,2}^*(z_j,t)\\
+ i\omega_1 Q_1 a_2(z_j,t) D^*(z_j,t),
\end{multline}
\begin{multline}
\frac{1}{v}\frac{\partial a_2(z_j,t)}{\partial t}  = -\frac{1}{2}\alpha a_2(z_j,t) + \frac{1}{4}g_2 \Gamma a_1(z_j,t) I_{1,2}(z_j,t)\\
- i\omega_2 Q_2 a_1(z_j,t) D(z_j,t),
\end{multline}
where the interaction integral $I_{1,2}(z_j,t)$ is computed as
\begin{multline}\label{eq:I_12}
    I_{1,2}(z_j,t_n) = \frac{\Delta t}{2}e^{-\frac{\Gamma}{2} n \Delta t}\bigg[ I_{1,2}(z_j,t_{n-1})\ \\
    +a_1^*(z_j,t_{n-1}) a_2(z_j,t_{n-1}) e^{\frac{\Gamma}{2}(n-1)\Delta t}\ \\ +a_1^*(z_j,t_{n}) a_2(z_j,t_{n}) e^{\frac{\Gamma}{2}n\Delta t} \bigg] ~.
\end{multline}
To integrate the envelope fields $a_1$ and $a_2$ in time, we use an Euler-Mayurama scheme~\cite{wang2013}, which yields the following finite-difference equations
\begin{multline}
    \label{eq:a1_up} a_1(z_j,t_{n+1}) = \left[1 - \frac{v\alpha \Delta t}{2}\right] a_1(z_j,t_n) \\
    -v\Delta t\left[\frac{g_1\Gamma I_{1,2}^*(z_j,t_n)}{4} - i\omega_1 Q_1 D^{*}(z_j,t_n)\right]a_2(z_j,t_n),
\end{multline}
\begin{multline}
    \label{eq:a2_up} a_2(z_j,t_{n+1}) = \left[1 - \frac{v\alpha \Delta t}{2}\right] a_2(z_j,t_n) \\
    +v\Delta t\left[\frac{g_2\Gamma I_{1,2}(z_j,t_n)}{4} - i\omega_2 Q_2 D(z_j,t_n)\right]a_1(z_j,t_n).
\end{multline}
The acoustic field is computed at each $z_j$ and $t_{n+1}$ after computing $a_{1,2}$, via the equation
\begin{equation}\label{eq:b_up}
        b(z_j,t_{n+1}) = i v_a \Omega Q_a I_{12}(z_j,t_{n+1}) + \sqrt{\Delta z}D(z_j,t_{n+1}).
\end{equation}
The $\sqrt{\Delta z}$ factor in front of $D(z_j,t_{n+1})$ ensures that the variance of $b$ is independent of the numerical grid resolution. 

Once all the fields are computed at $t_{n+1}$, we repeat the drift steps in~\eqref{eq:A1shift} and~\eqref{eq:A2shift} and the entire process is iterated until the optical fields have propagated across the waveguide. The steps of this numerical method are given in Algorithm~\ref{alg:num_alg}.

\begin{algorithm}
\caption{Numerical algorithm}\label{alg:num_alg}
\begin{algorithmic}[1]
\State Compute $D(z_j,t_n)$ for all $t_n$
\State Compute $\phi_{1,2}(t_n)$ for all $t_n$
\State Set $a_{1,2}=0$ inside $z\in [0,L]$
\For{$n = 1$ to $N_t-1$}\Comment{$N_t=$ size of time grid}
\State Insert noisy boundary conditions in $a_{1,2}$ at $t_n$
\State Shift optical fields $a_{1,2}$ in space by $\Delta z$
\State Compute interaction integral $I_{1,2}(z_j,t_n)$ 
\State Compute $a_{1,2}(z_j,t_{n+1})$ from $a_{1,2}(z_j,t_n)$
\State Compute $b(z_j,t_{n+1})$
\EndFor{\textbf{end for}}
\end{algorithmic}
\end{algorithm}

%%% --- Statistical Properties of the Fields --- %%%
\subsection{Statistical properties of the fields}
The iterative scheme in Algorithm~\ref{alg:num_alg} computes a single realization of the SBS interaction given a specific set of input parameters. We must repeat this process $M$ times with the same input parameters to build an ensemble of $M$ independent simulations, from which statistical properties may be calculated. For instance, the true average of the power for all three fields ($P_{1,2}$ for the optical fields and $P_a$ for the acoustic field) may be calculated as
\newcommand{\q}{1,2}
\begin{equation}
    \left\langle P_{\q}(z_j,t_n)\right\rangle = \left\langle \left| a_{\q}(z_j,t_n)\right|^2\right\rangle
    \approx \frac{1}{M}\sum_{m = 1}^{M} \left| a_{\q}^{(m)}(z_j,t_n)\right|^2,
\end{equation}
\begin{equation}
    \left\langle P_a(z_j,t_n)\right\rangle = \left\langle \left| b(z_j,t_n)\right|^2 \right\rangle \\
    \approx \frac{1}{M}\sum_{m = 1}^{M} \left| b^{(m)}(z_j,t_n)\right|^2,
\end{equation}
where $m$ refers to a specific realization of each process. Similarly, we compute the standard deviation in the power at each point $(z_j,t_n)$ as
\begin{align}
    \text{std}\left[P_{\q}(z_j,t_n)\right] &= \sqrt{ \left\langle \left[P_{\q}(z_j,t_n)\right]^2 \right\rangle - \left\langle P_{\q}(z_j,t_n) \right\rangle^2 }, \\
    \text{std}\left[P_{a}(z_j,t_n)\right] &= \sqrt{ \left\langle \left[P_{a}(z_j,t_n)\right]^2 \right\rangle - \left\langle P_{a}(z_j,t_n) \right\rangle^2 }.
\end{align}
The standard deviation is useful when comparing with experiments, since it gives a quantitative measure of the size of the power fluctuations in the measured optical fields.

%%%%%%%%%%%%%%%%%%%%%%%%%%%%%%%%%%%%%%%%%%%%%%%%%%%%
% Results and Discussion (UPDATED MARCH 17th 2021)
%%%%%%%%%%%%%%%%%%%%%%%%%%%%%%%%%%%%%%%%%%%%%%%%%%%%
\section{Results and Discussion}
We demonstrate the numerical method by simulating the SBS interaction of the three fields with both thermal noise ($T=300$ K, $\Delta\nu_B = 30$ MHz) and laser noise ($\Delta\nu_L = 100$ kHz), using a chalcogenide waveguide of length 50 cm, with the properties in Table~\ref{table:tab1}. Although our formalism includes optical loss through the factor $\alpha$, we have chosen $\alpha = 0$ in the simulations to focus on the effect of net SBS gain and pulse properties on the noise. Here we study the noisy SBS interaction in two different cases: spontaneous scattering and stimulated scattering, and investigate the effects of pump width and SBS gain on the noise properties of the Stokes field.

% TABLE 1
\begin{table}[h]
\centering
\begin{tabular}{||c c||} 
 \hline
Parameter & Value \\ [0.5ex] 
 \hline\hline
Waveguide length $L$ & 50 cm \\
Waveguide temperature $T$ & 300 K \\
Refractive index $n$ & 2.44 \\
Acoustic velocity $v_a$ & 2500 m/s \\
Brillouin linewidth $\Delta\nu_B$ & 30 MHz\\
Brillouin shift $\Omega/2\pi$ & 7.7 GHz \\
Brillouin gain parameter $g_0$ & 423 m$^{-1}$W$^{-1}$\\
Optical wavelength $\lambda$ & 1550 nm \\
Laser linewidth $\Delta\nu_L$ & 100 kHz \\
Peak pump power & 1 W \\
Peak Stokes power & 0$-$1 mW\\
Simulation time $t_f$ & up to 80 ns\\
Pump pulse FWHM & 0.5$-$5 ns \\
Stokes pulse FWHM & 1 ns \\
Grid size (space) $N_z$ & 1001\\
Grid size (time) $N_t$ & 2601\\ 
Step-size $\Delta t$ & 4.07 ps\\ [1ex]
 \hline
\end{tabular}
\caption{Simulation parameters using a chalcogenide waveguide of the type shown in~\cite{xie2019}.}
\label{table:tab1}
\end{table}

\subsection{The spontaneous Brillouin scattering case}
\begin{figure}[H]
\centering
\includegraphics[width=0.5\textwidth]{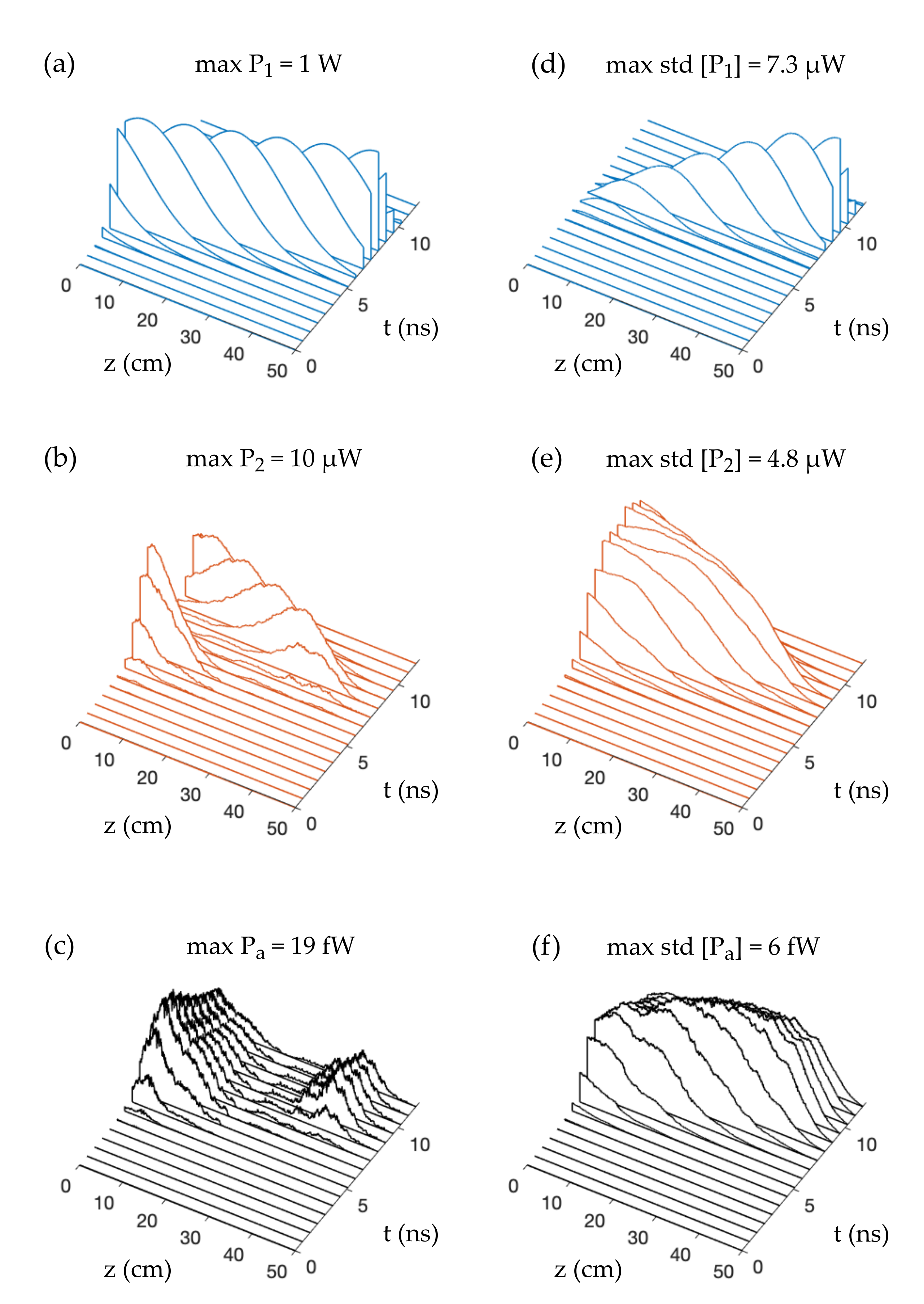}
\caption{Waterfall plots for a single numerical realization of (a) pump power, (b) Stokes power and (c) acoustic power in the spontaneous scattering case, using a Gaussian pump of FWHM 2 ns and peak power of 1 W. Plots (d)$-$(f) show the standard deviation of the field powers at each point $(z,t)$, calculated from 100 independent realizations of the SBS interaction.}
\label{fig:WaterfallPlots}
\end{figure}

We first consider the situation in which there is no input Stokes field from an external laser source, and the Stokes arises purely from the interaction between the pump and the thermal field --- this situation is customarily referred to as {\em spontaneous} or {\em spontaneously-seeded} Brillouin scattering. We specify a Gaussian pump pulse of varying widths and constant peak power, with input phase noise ($\Delta \nu_L = 100$ kHz). Setting the waveguide temperature at 300 K and the pump FWHM of 2 ns, in Fig.~\ref{fig:WaterfallPlots}(a)$-$(c) we see that the thermal acoustic field interacts with the pump to generate an output Stokes signal. At the same time, the Stokes field depletes some of the pump and amplifies the acoustic field, which leads to more Stokes energy being generated. The noisy character of the Stokes field in Fig.~\ref{fig:WaterfallPlots}(b) is due to the incoherent thermal acoustic background, which generates multiple random Stokes frequencies. In this short-pump regime, the SBS amplification is small, and the generated Stokes field remains incoherent. 

% --- FIG. Waterfall Plots for Spontaneous Scattering (2 ns pump)

\begin{figure}[H]
\centering
\includegraphics[width=0.5\textwidth]{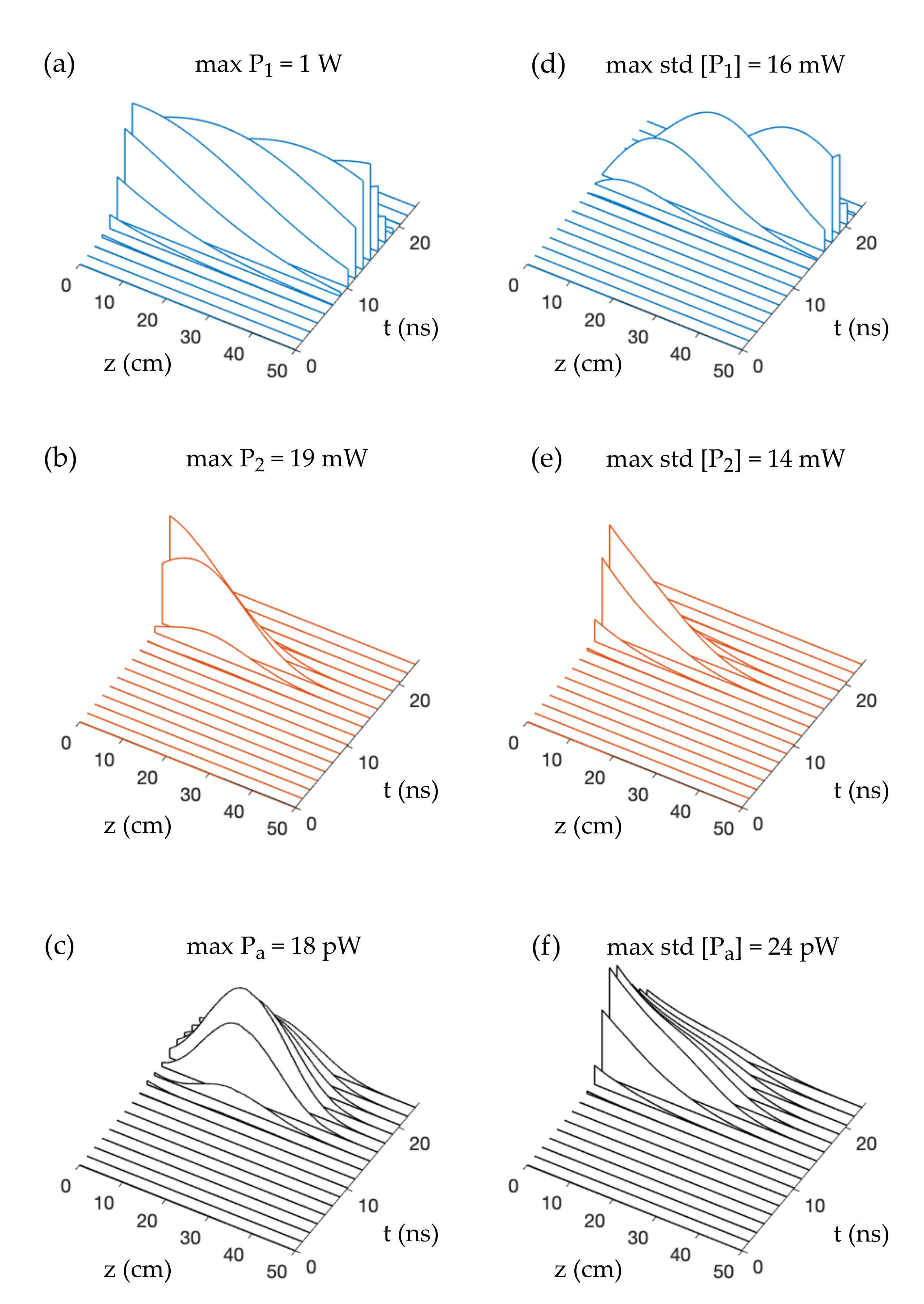}
\caption{Waterfall plots for a single numerical realization of (a) pump power, (b) Stokes power and (c) acoustic power in the spontaneous scattering case, using a Gaussian pump of FWHM 5 ns and peak power of 1 W. Plots (d)$-$(f) show the standard deviation of the field powers at each point $(z,t)$, calculated from 100 independent realizations of the SBS interaction.}
\label{fig:WaterfallPlots2}
\end{figure}

\begin{figure}[H]
\centering
\includegraphics[width=0.5\textwidth]{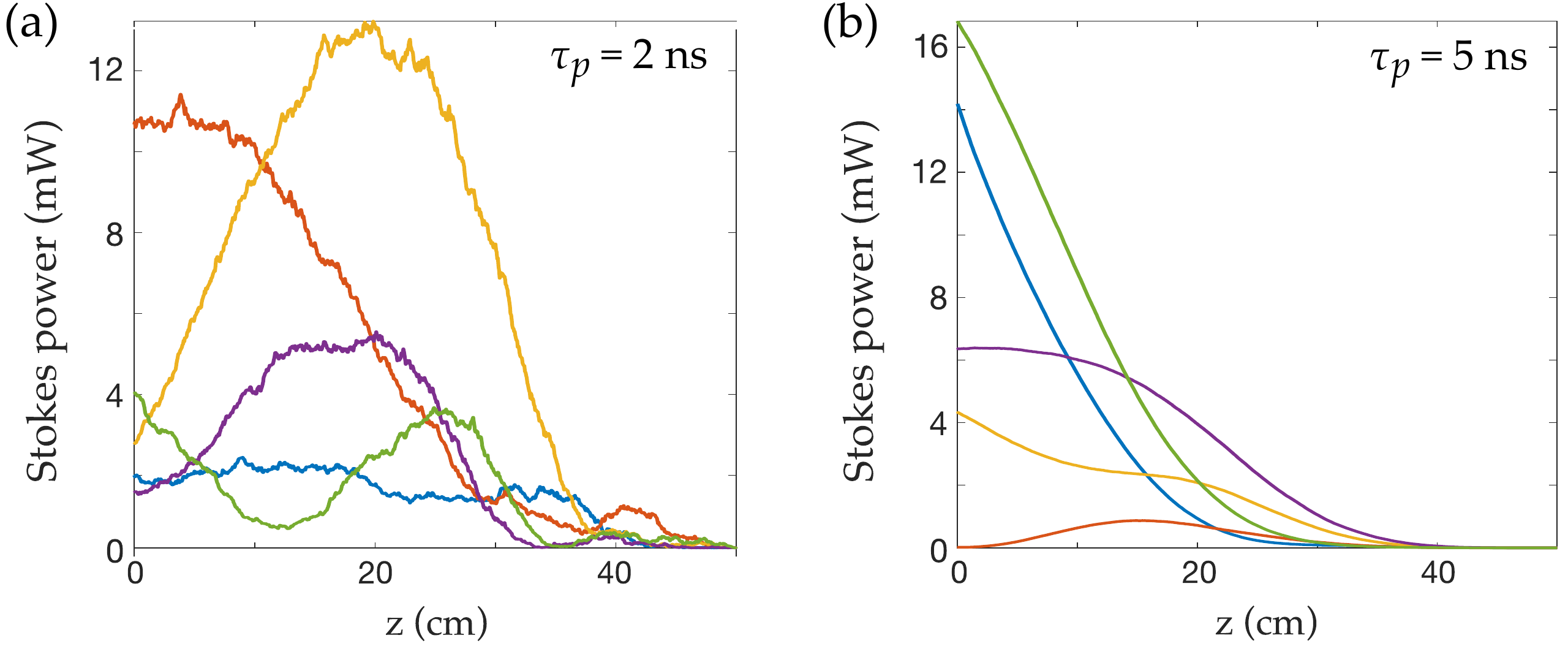}
\caption{Multiple independent realizations of the spontaneously generated Stokes power across the waveguide for (a) 2 ns wide pump and (b) 5 ns wide pump. These snapshots are taken at the time when the peak of the pump pulse reaches $z=50$ cm.}
\label{fig:Spontaneous_Multiple}
\end{figure}

As we increase the width of the pump to 5 ns, the net SBS gain in the waveguide also increases. In this long-pump regime, the (spontaneously-generated) Stokes field is amplified coherently, as shown in Fig.~\ref{fig:WaterfallPlots2}(b). However, it should be noted that, although the Stokes output becomes smooth, there is significant variation in the 
peak Stokes power from one independent realization to the next, as illustrated in Fig.~\ref{fig:Spontaneous_Multiple}(a) and (b).
The standard deviation of the Stokes power over multiple independent realizations increases with longer pump pulses, as shown in Fig.~\ref{fig:WaterfallPlots2}(e).

% --- FIG. Waterfall Plots for Spontaneous Scattering (5 ns Pump)

% --- FIG. Waterfall Plots for Spontaneous Scattering Multiple

\begin{figure}[H]
\centering
\includegraphics[width=0.5\textwidth]{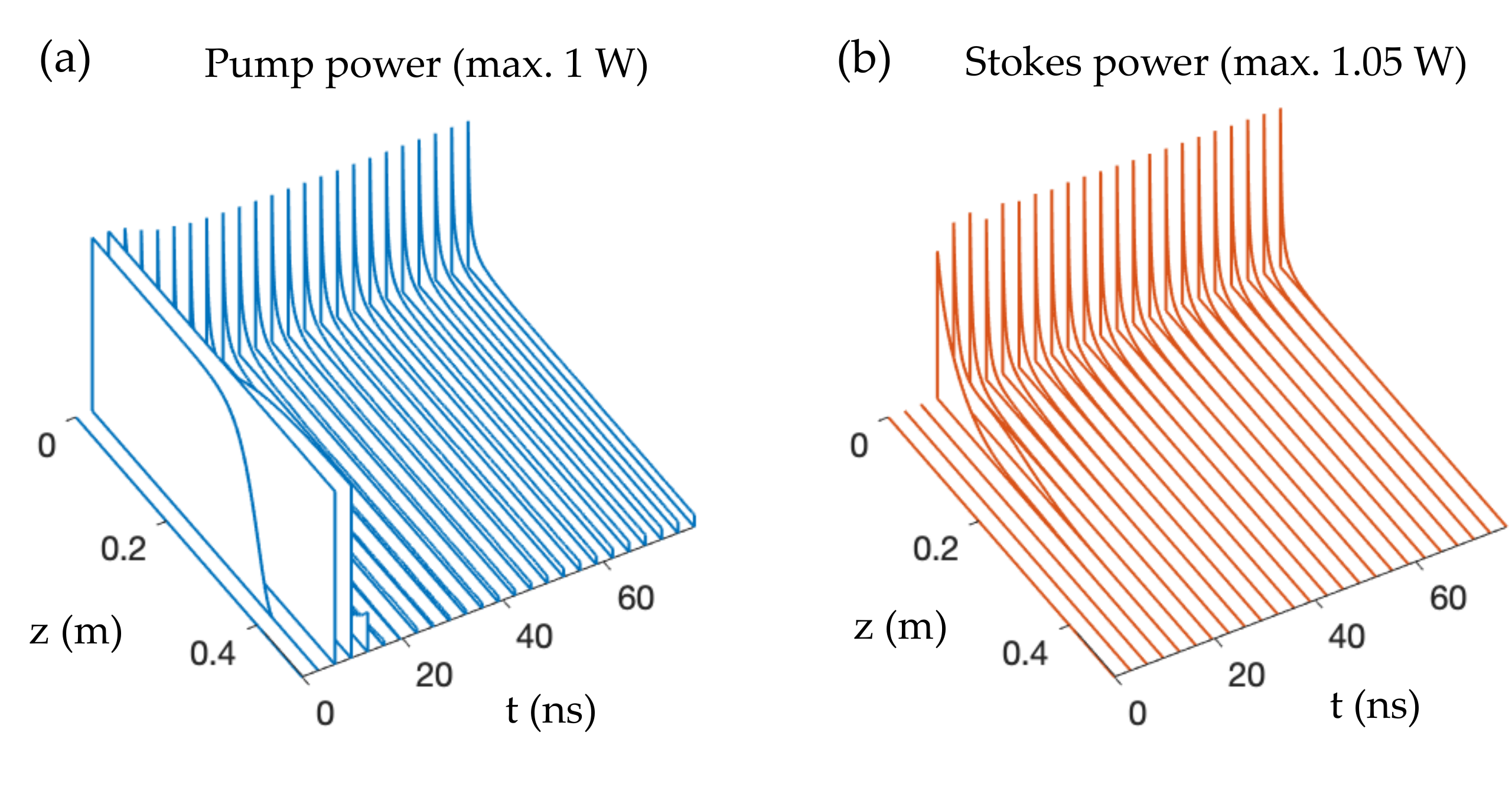}
\caption{Waterfall plots for a single numerical realization of (a) pump power and (b) Stokes power in the spontaneous scattering case, using a CW pump with 1 W peak power.}
\label{fig:WaterfallPlots3}
\end{figure}

As the pump becomes very long we approach the CW regime, in
which the pump power ramps up quickly at $z=0$ and is kept at a constant value. If the waveguide is sufficiently long, the spontaneously generated Stokes field is amplified coherently until pump depletion begins to take effect, initially at $z=0$ and then throughout the length of the waveguide,
until both Stokes and pump fields relax into the steady-state configuration in which the pump decreases exponentially, as shown in Fig.~\ref{fig:WaterfallPlots3}(a)$-$(b).
When such a steady state is reached, the depletion induced by the spontaneously-seeded Stokes may inhibit Brillouin scattering
from an input Stokes pulse injected at $z=L$.

% --- FIG. Waterfall Plots for Spontaneous Scattering (CW pump)

% --- FIG. SBS threshold plots
\begin{figure}[ht]
\centering
\includegraphics[width=0.5\textwidth]{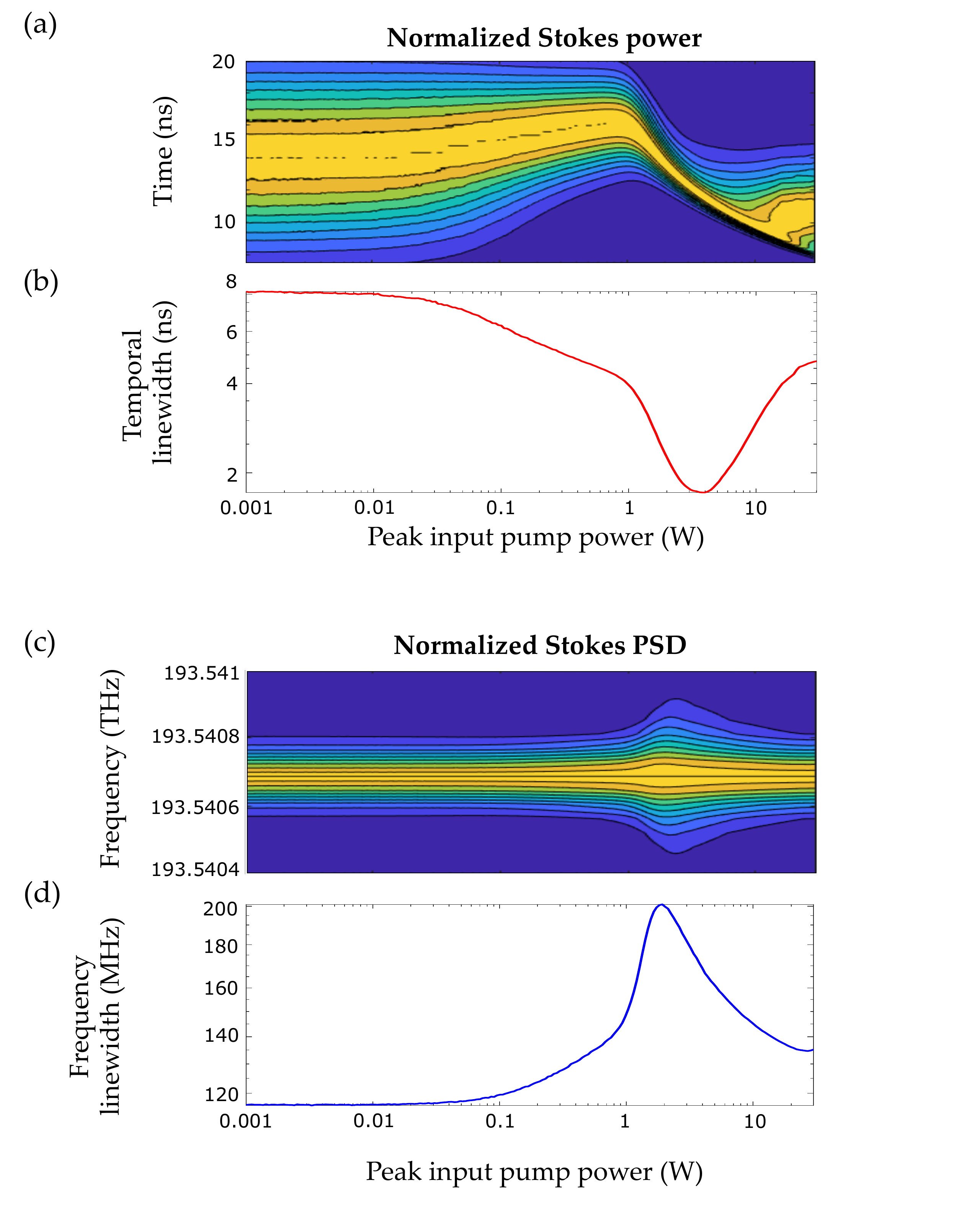}
\caption{Computations of the spontaneously generated Stokes field at $z=0$ over 500 independent realizations, using a 5 ns Gaussian pump pulse, with varying input peak pump power. Plot (a) shows the output ensemble averaged Stokes power at $z=0$, normalized by the maximum power at each input pump, (b) shows the FWHM of the Stokes in time domain. Plot (c) shows the normalized power spectral density (PSD) of the Stokes field, and (d) is the FWHM of the Stokes in frequency domain.}
\label{fig:Stokes_threshold}
\end{figure}

Returning to the pulsed case, we investigate the effect of increasing the peak pump power, and therefore the overall SBS gain, on the amplification of the spontaneous Stokes field. 
Figure~\ref{fig:Stokes_threshold} shows how the Stokes spectral linewidth increases for input pump powers between 0.1$-$2 W
for a Gaussian pump pulse with fixed FWHM of 5 ns.
The increase in linewidth occurs due to the transition from linear to nonlinear SBS amplification: in the linear amplification regime, the spontaneously generated Stokes field retains a constant temporal width while its peak power increases with input pump power. In the nonlinear amplification regime, the Stokes field undergoes temporal compression as a result of the central peak of the pulse being amplified faster than the tails. Beyond 2 W of peak pump power, the spectral linewidth of the Stokes field narrows as pump depletion becomes significant, because the Stokes field is prevented from uniformly experiencing exponential gain throughout the waveguide, an effect which is also observed in the CW pump case~\cite{gaeta1991stochastic}. 

\subsection{The effect of laser phase noise}
\begin{figure}[h]
\centering
\includegraphics[width=0.5\textwidth]{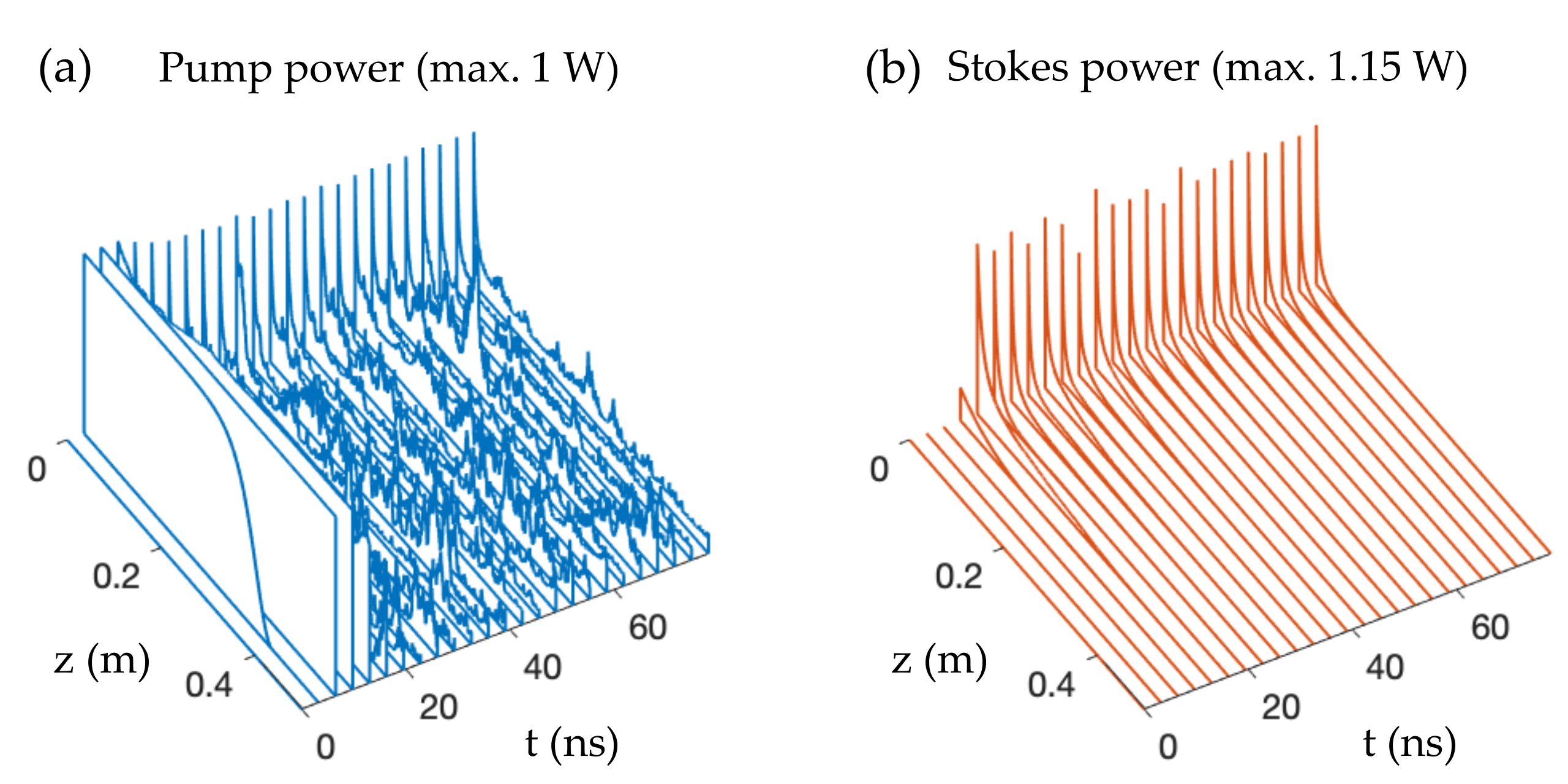}
\caption{Waterfall plots for a single numerical realization of (a) pump power and (b) Stokes power in the spontaneous scattering case, using a CW pump with 1 W peak power and a laser linewidth of 100 MHz.}
\label{fig:WaterfallPlots4}
\end{figure}

Our previous simulations included laser phase noise corresponding to a laser linewidth of 100 kHz in the pump. This is equivalent to a coherence time of $\tau_{\text{coh}} = 3.2$ $\mu$s, which is at least 100 times larger than the characteristic time of the SBS interaction in Fig.~\ref{fig:WaterfallPlots}$-$\ref{fig:WaterfallPlots3}. For this reason it is understandable that no contribution from the laser phase noise to the optical or acoustic fields was observed. The contribution of laser phase noise can however be observed if the linewidth of the pump is suffiently broad. We therefore consider the CW-pump regime with zero Stokes input power, with a laser linewidth of 100 MHz, which corresponds to a coherence time of 3.2 ns (Fig.~\ref{fig:WaterfallPlots4}). We see a significant contribution from the laser phase noise in the form of amplitude fluctuations,
which are completely absent in the 100 kHz linewidth case (Fig.~\ref{fig:WaterfallPlots3}).
From this we infer that when the laser coherence time $\tau_{\text{coh}}$ is comparable to the pulse widths $\tau_{p,s}$, the fluctuations in the phase are fast enough to be transferred to the envelope of the pulse. However, when $\tau_{\text{coh}} \gg \tau_{p,s}$, the noisy character of the envelope fields will vanish. This has important implications for the case of pulsed SBS: phase noise can only play a significant role in the interaction if $\tau_{\text{coh}} \leq \tau_{p,s}$. For lasers with a relatively small linewidth, such as in the kHz range, phase noise will only become a significant effect when operating in the long-pulse or CW regime. 

% --- FIG. Waterfall Plots for Spontaneous Scattering (CW pump Phase Noise 100 MHz)

\subsection{The stimulated Brillouin scattering case}
\begin{figure}[h]
\centering
\includegraphics[width=0.5\textwidth]{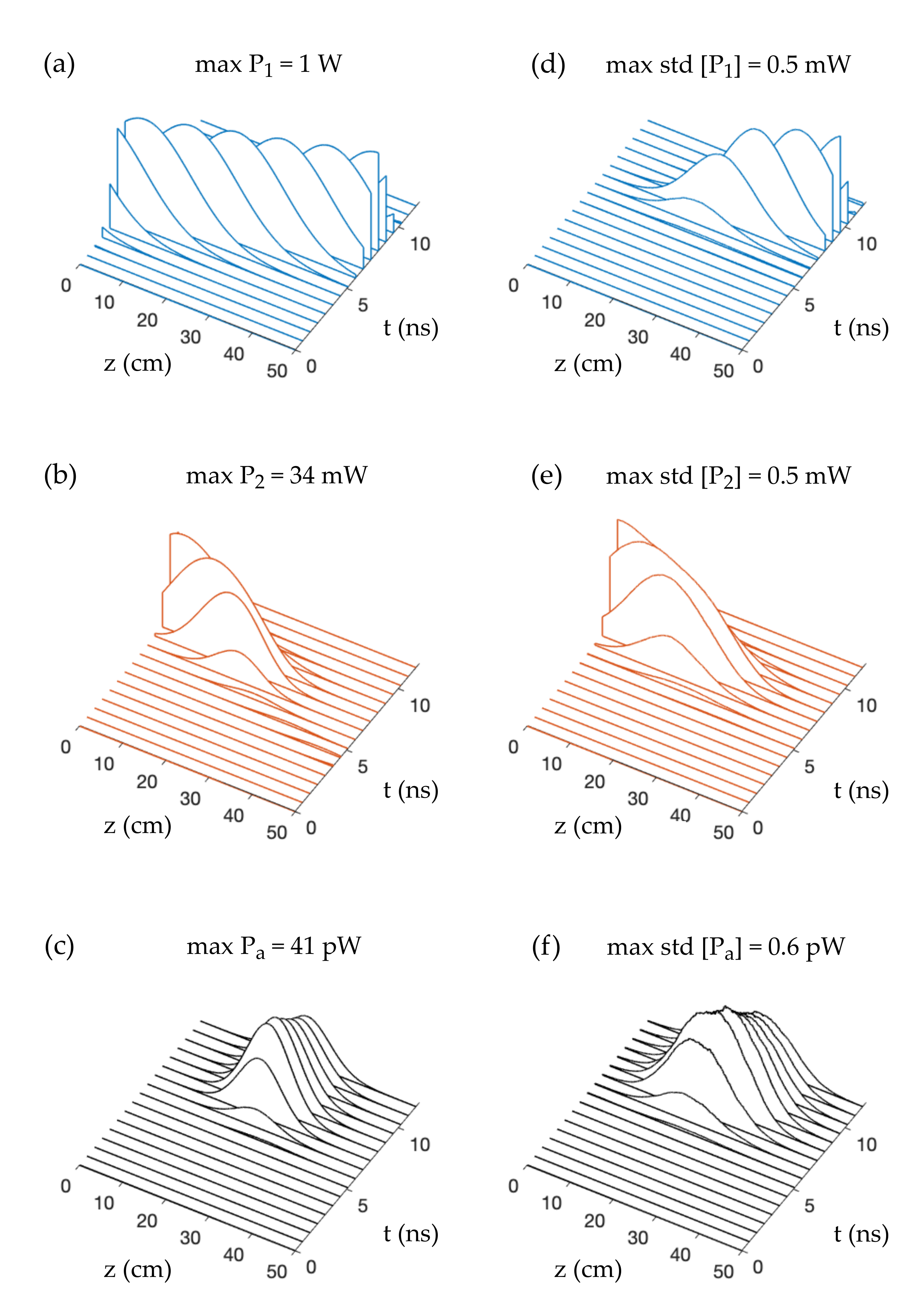}
\caption{Waterfall plots for a single numerical realization of (a) pump power, (b) Stokes power and (c) acoustic power in the stimulated scattering case, using a Gaussian pump pulse of width 2 ns and peak power 1 W. The input Stokes pulse has width 1 ns and peak power 1 mW. Plots (d)$-$(f) show the standard deviation in the fields at each point $(z,t)$ for 100 independent realizations of the SBS interaction.}
\label{fig:WaterfallPlots5}
\end{figure}

\begin{figure}[h]
\centering
\includegraphics[width=0.5\textwidth]{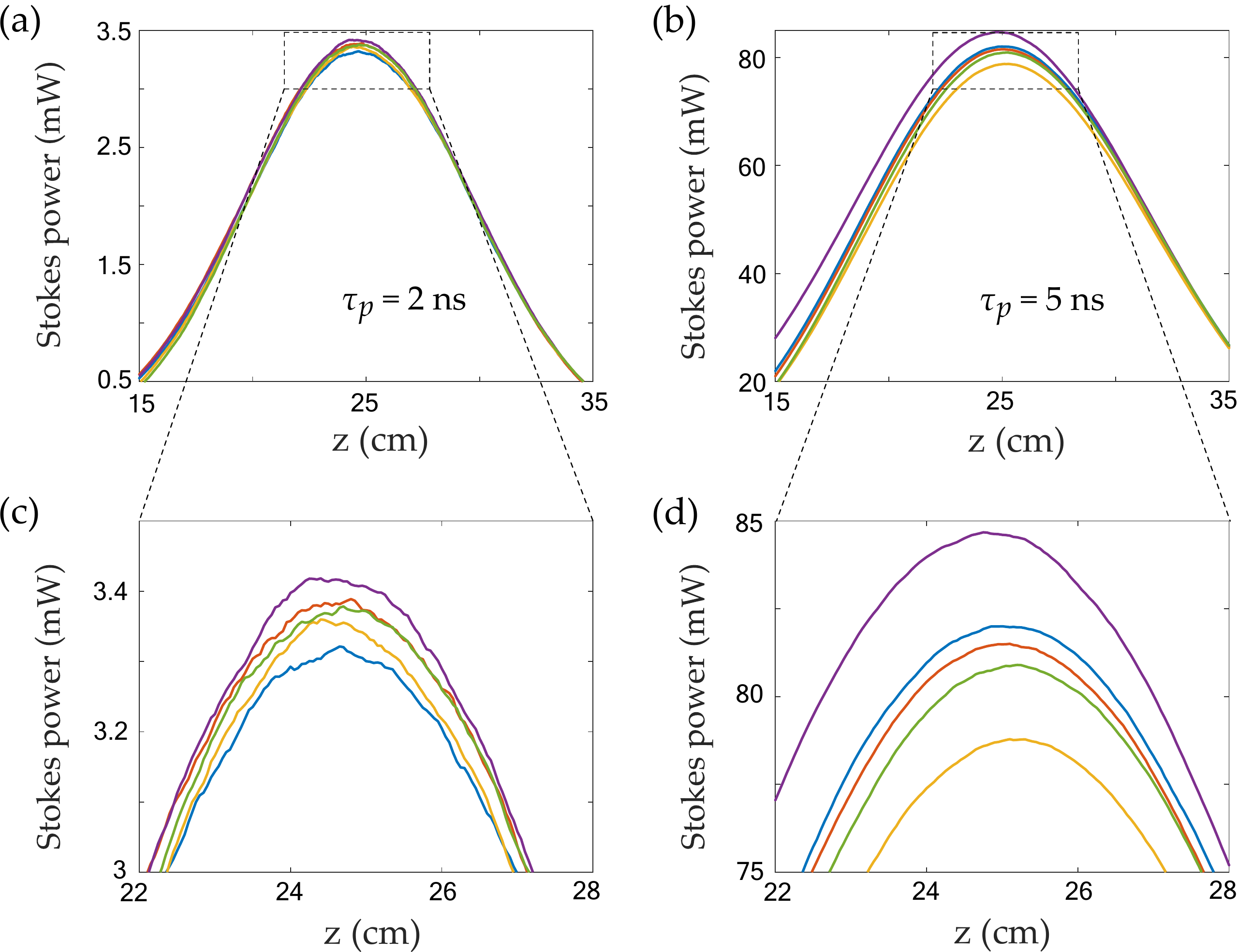}
\caption{Multiple independent realizations of the Stokes power across the waveguide for (a) 2 ns wide pump and (b) 5 ns wide pump. These snapshots are taken at the time in which the peaks of the pump and Stokes meet in the middle of the waveguide.}
\label{fig:Stokes_plots_1}
\end{figure}

We now examine the case of \textit{seeded} Brillouin scattering, in
which a Stokes signal is injected at $z = L$. We first consider a 1 mW peak power Stokes pulse of FWHM 1 ns in the same chalcogenide waveguide as before. The pump is a Gaussian pulse of constant peak power of 1 W, with a width of 2 ns. As can be seen in Fig.~\ref{fig:WaterfallPlots5}, the Stokes pulse remains smooth throughout the interaction, and although the standard deviation over 100 independent realizations is approximately 1.4\% of the peak value, there are no visible fluctuations in the power across space or time in Fig.~\ref{fig:WaterfallPlots5}(b). A closer look at multiple individual realizations in Fig.~\ref{fig:Stokes_plots_1}(a) reveals that there is a measurable level of variation in the Stokes power, although each individual realization of the Stokes field is smooth. By increasing the pump width to 5 ns as shown in Fig.~\ref{fig:Stokes_plots_1}(b), we also increase the standard deviation in the Stokes, however each independent realization appears smoother compared to Fig.~\ref{fig:Stokes_plots_1}(a). This further demonstrates how in the longer pump, high SBS gain regime, the amplification of the Stokes is sufficient to cancel random phase differences in the Stokes field, as we observed in the spontaneous scattering case in Fig.~\ref{fig:WaterfallPlots2}.

% --- FIG. Waterfall Plots for Stimulated Scattering (5 ns pump case)

% --- FIG. Stokes Plots at fixed time for (a) 2 ns pump and (b) 5 ns pump

\subsection{Convergence of the method}
We now study the convergence of the numerical method by looking at the statistical properties of the power in each field at fixed points on $(z,t)$. We use a default minimum step-size in time $\Delta t_{\text{min}} = 40.7$ fs against which we compare the results for larger step-sizes $\Delta t$. We compute the relative error in the power and variance of the power, taken over 1,000 independent realizations. These results correspond to what is known as weak convergence in stochastic differential equations~\cite{kloeden1992}, where the mean value of a random quantity, in our case the power, converges at a specific rate with respect to the step-size used.

\begin{figure}[ht]
\centering
\includegraphics[width=0.5\textwidth]{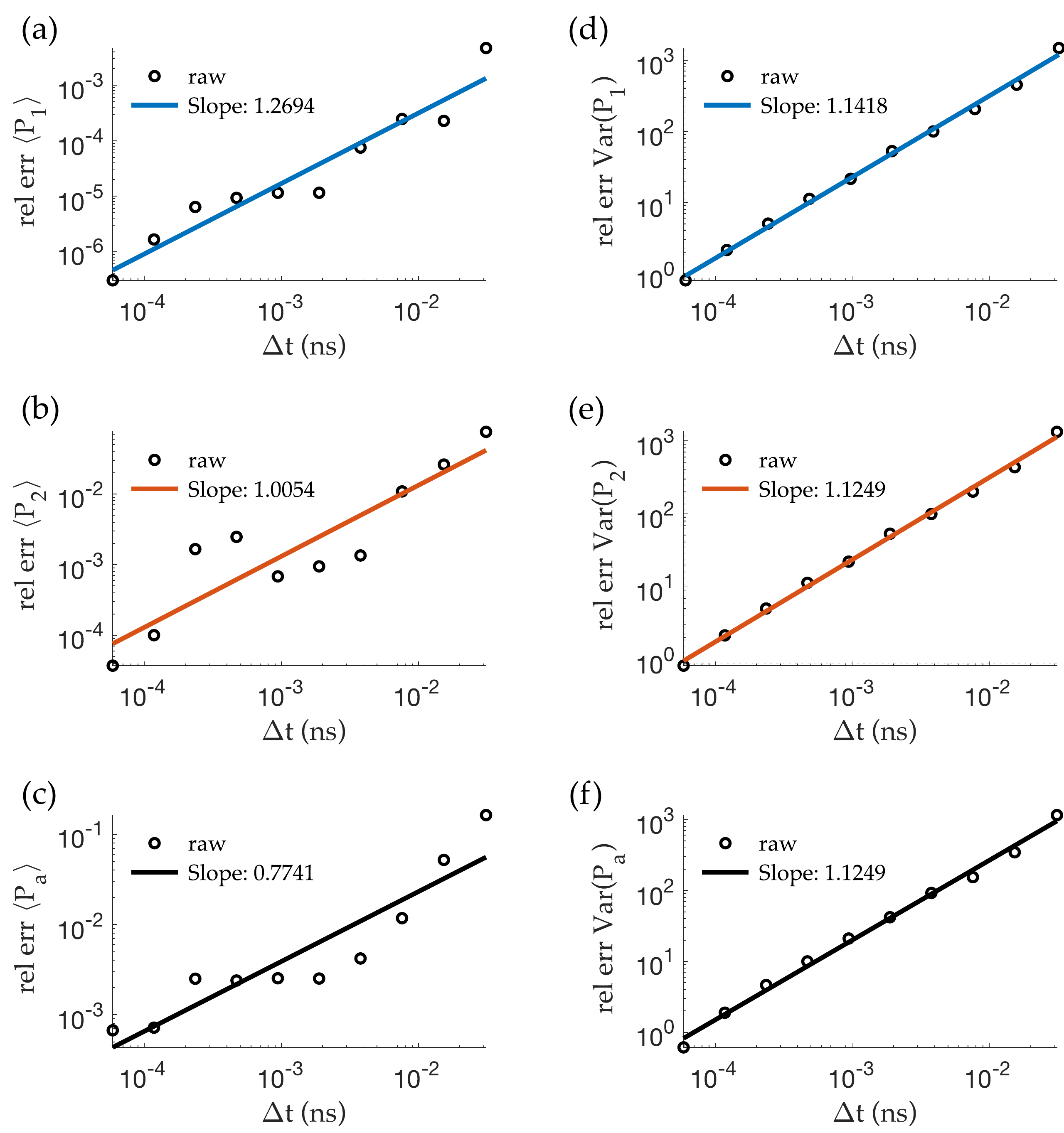}
\caption{Convergence plots showing the relative error in the ensemble averaged powers (a)$-$(c) and in the variance of the powers (d)$-$(f), as a function of the step-size $\Delta t$ used in the numerical grid. The reference step-size used is $\Delta t_{\text{min}} = 40.7$ fs. The calculations are based on a sample size of 1,000 independent simulations of the fields. The test problem consists of two optical Gaussian pulses for the pump and Stokes of width 1 ns, with peak powers pump 100 mW (pump) and 10 $\mu$W (Stokes). The statistical properties of $P_1$, $P_2$ and $P_a$ are calculated from $P_1(L,t_{\text{max}})$, $P_2(0,t_{\text{max}})$ and $P_a(L/2,t_{\text{max}})$ respectively, where $t_{\text{max}}$ is the time at which the peaks of the optical pulses reach the opposite ends of the waveguide. The computations include thermal noise in the waveguide at temperature 300 K, and input laser phase noise with linewidth 100 kHz. The waveguide properties are given in Table~\ref{table:tab1}.}
\label{fig:Convergence}
\end{figure}

The results for the convergence computations are shown in Fig.~\ref{fig:Convergence}. As expected from the Euler-Mayurama scheme~\cite{kloeden1992}, the convergence rate is at most linear for the mean power of all three fields. A similar rate of convergence is recorded for the variance in each power, showing a one-to-one error reduction with step-size. Although some higher order methods exist which implement higher order Taylor expansions and Runge-Kutta schemes~\cite{kloeden1992,honeycutt1992,honeycutt1992stochastic,tocino2002}, these methods only work with ordinary stochastic differential equations; numerical methods for partial stochastic differential equations are an active area of research in applied mathematics~\cite{zhang2017num}.

% --- FIG 3 - Convergence plots

%%%%%%%%%%%%%%%%%%%%%%%%%%%%%%%%%%%%%%%%%%%%%%%%%%%%
% CONCLUSIONS
%%%%%%%%%%%%%%%%%%%%%%%%%%%%%%%%%%%%%%%%%%%%%%%%%%%%
\section{Conclusion}
We have presented a numerical method by which the fully-dynamic coupled SBS equations in both CW and pulsed scenarios with thermal and laser noise can be solved. The method offers linear convergence in both the average power and variance of the power of the optical and acoustic fields, with variances that do not depend on step-size. From our simulations, we find that the noise properties of the fields rely on the length of the optical pulses involved as well as on the net SBS gain in the waveguide. For short-pump, low gain regimes, the spontaneous Stokes field is incoherently amplified and exhibits large spatial and temporal fluctuations, whereas for the long-pump, high gain regime the field is amplified coherently, resulting in a smooth field but with large variations in peak power between independent realizations. Similar observations are made for the stimulated scattering case using a Stokes signal. We also find that laser phase noise does not play a significant role in the SBS interaction unless the laser coherence time is comparable to the characteristic time-scales of the SBS interaction.

%%%%%%%%%%%%%%%%%%%%%%%%%%%%%%%%%%%%%%%%%%%%%%%%%%%%
% APPENDICES
%%%%%%%%%%%%%%%%%%%%%%%%%%%%%%%%%%%%%%%%%%%%%%%%%%%%
\appendix
\section{Appendix A}
The integral term in~\eqref{eq:D_analytic} can be evaluated using the properties of It\^{o} integrals. Firstly, since the integrand is a deterministic function of time, and $\da{W_j(s)}$ is a normally distributed stochastic process, the integral is also a normally distributed stochastic process. Secondly, $\da{W_j(s)}$ is a complex-valued process, so the integral can be split into two statistically independent real-valued integrals
\begin{multline}
    \int_{0}^{t} e^{-\frac{\Gamma}{2}(t-s)} \da{W_j(s)} = \frac{1}{\sqrt{2}} \int_{0}^{t} e^{-\frac{\Gamma}{2}(t-s)} \da{W_j^{(1)}(s)} \\
    +i \frac{1}{\sqrt{2}} \int_{0}^{t} e^{-\frac{\Gamma}{2}(t-s)} \da{W_j^{(2)}(s)}, 
\end{multline}
each of these real integrals will have the same statistical properties, namely
\begin{equation}
    \left\langle \int_{0}^{t} e^{-\frac{\Gamma}{2}(t-s)} \da{W_j^{(q)}(s)} \right\rangle = 0.
\end{equation}
The variance is derived using the It\^{o} isometry property for a stochastic process $X(t)$~\cite{oksendal2003}
\begin{equation}
   \left\langle \left(\int_{0}^{t} X(s) \da{W(s)}\right)^2 \right\rangle = \left\langle\int_{0}^{t} X^2(s) \da{s}\right\rangle.
\end{equation}
Using this property, we write
\begin{equation}
    \left\langle \left(\int_{0}^{t} e^{-\frac{\Gamma}{2}(t-s)} \da{W_j^{(q)}(s)}\right)^2\right\rangle = \frac{1}{\Gamma}\left(1-e^{-\Gamma t}\right),
\end{equation}
which leads to the result for the variance
\begin{equation}
    \text{Var}\left[\int_{0}^{t} e^{-\frac{\Gamma}{2}(t-s)} \da{W_j^{(q)}(s)} \right] = \frac{1}{\Gamma}\left(1-e^{-\Gamma t}\right).
\end{equation}
This means the integral can be computed as a normal random variable as
\begin{equation}
    \int_{0}^{t} e^{-\frac{\Gamma}{2}(t-s)} \da{W_j(s)} \sim \sqrt{\frac{1-e^{-\Gamma t}}{2\Gamma}}\left[\mathcal{N}_{z_j,t}^{(1)}(0,1) + i {N}_{z_j,t}^{(1)}(0,1)  \right],
\end{equation}
which leads to~\eqref{eq:D2}.

\section*{Disclosures}
The authors declare no conflicts of interest.

\section*{Acknowledgements}
The authors acknowledge funding from the Australian Research Council (ARC) (Discovery Project DP200101893), the Macquarie University Research Fellowship Scheme (MQRF0001036) and the UTS Australian Government Research Training Program Scholarship (00099F).  Part of the numerical calculations were performed on the UTS Interactive High Performance Computing (iHPC) facility. 

\section*{Data Availability Statement}
Data underlying the results presented in this paper are not publicly available at this time, but the data and accompanying code used to generate it can be obtained from the authors on reasonable request.

%%%%%%%%%%%%%%%%%%%%%%%%%%%%%%%%%%%%%%%%%%%%%%%%%%%%
% REFERENCES
%%%%%%%%%%%%%%%%%%%%%%%%%%%%%%%%%%%%%%%%%%%%%%%%%%%%
\bibliography{josab_references_2020}

\end{document}